\documentclass[letter,bibyear]{aa} 

\usepackage{epsfig}
\usepackage{amsmath}
\usepackage{subfigure}
\usepackage{natbib}
\usepackage{multirow}
\usepackage{color}
\usepackage{longtable}
\usepackage{graphicx}
\usepackage{epstopdf}
\usepackage{booktabs}
\usepackage{txfonts}

\newcommand{\jmst}{J.~Mol.~Struct.}

\newcommand{\kms}{km s$^{-1}$}

\begin{document}

\title{Discovery of HCCCH$_2$CCH in TMC-1 with the QUIJOTE line survey\thanks{Based on 
observations carried out
with the Yebes 40m telescope (projects 19A003,
20A014, 20D023, 21A011, 21D005, and 23A024). The 40m
radio telescope at Yebes Observatory is operated by the Spanish Geographic Institute
(IGN, Ministerio de Transportes, Movilidad y Agenda Urbana).}}

\author{
R.~Fuentetaja\inst{1},
M.~Ag\'undez\inst{1},
C.~Cabezas\inst{1},
B.~Tercero\inst{2,3},
N.~Marcelino\inst{2,3},
J.~R.~Pardo\inst{1},
P.~de~Vicente\inst{2},
J.~Cernicharo\inst{1}
}

\institute{Dept. de Astrof\'isica Molecular, Instituto de F\'isica Fundamental (IFF-CSIC),
C/ Serrano 121, 28006 Madrid, Spain. \newline \email r.fuentetaja@csic.es, jose.cernicharo@csic.es
\and Centro de Desarrollos Tecnol\'ogicos, Observatorio de Yebes (IGN), 19141 Yebes, Guadalajara, Spain.
\and Observatorio Astron\'omico Nacional (OAN, IGN), C/ Alfonso XII, 3, 28014, Madrid, Spain.
}

\date{Received; accepted}

\abstract{We present the first detection in space of 1,4-pentadiyne. It has
been found towards TMC-1 with the QUIJOTE line survey in the 31-50 GHz range. 
We observed a total of 17 transitions with $J$ = 2 up to 13 and $K_a$ = 
0,1 and 2. The observed transitions allowed us to derive a rotational temperature 
of 9.5 $\pm$ 0.5 K and a column density of (5.0 $\pm$ 0.5)$\times$ 10$^{12}$ cm$^{-2}$. 
This molecule was the last non-cyclic isomer of the C$_5$H$_4$ family that could be detected via radio astronomy. A computational chemistry study was performed to determine the energies of the five most stable isomers. The isomer ($c$-C$_3$H$_3$CCH) has a considerably higher energy than the 
others, and it has not yet been detected.
To better understand the chemical reactions involving these species, 
we compared the ethynyl and cyano derivatives. The observed 
abundances of these species are in good agreement with the branching ratios of 
the formation reactions studied with our chemical model of TMC-1.
}

\titlerunning{HCCCH$_2$CCH in TMC-1}
\authorrunning{Fuentetaja et al.}

\maketitle

\section{Introduction}

Studies of the TMC-1 dark cloud have greatly improved our understanding of the chemistry 
of the interstellar medium. Its carbon-rich composition allows the formation of 
a large variety of species. A large number of new molecules have been detected in 
recent years with the Yebes 40m radio telescope via the QUIJOTE\footnote{Q-band Ultrasensitive Inspection Journey to the Obscure TMC-1 Environment} line 
survey \citep[][]{Cernicharo2021a} and with the Green Bank 100m telescope via the GOTHAM\footnote{GBT Observations of TMC-1: Hunting Aromatic Molecules} line survey  
\citep[][]{McGuire2018}. 
Recent discoveries include several cyclic molecules, such as benzyne, cyclopentadiene, 
indene, ortho-benzyne, and fulvenallene \citep{Cernicharo2021a, Cernicharo2021b, Cernicharo2022}. Other cycles detected in TMC-1 are the cyano and ethynyl derivatives of cyclopentadiene \citep{McCarthy2021, Lee2021,Cernicharo2021c} and benzene \citep{Loru2023,McGuire2018}, and cyano derivatives of naphthalene and indene \citep{McGuire2021, Sita2022}. Among the long carbon chains, molecules such as vinyl acetylene \citep{Cernicharo2021d}, allenyl acetylene \citep{Cernicharo2021e}, butadiynylallene \citep{Fuentetaja2022}, ethynylbutatrienylidene \citep{Fuentetaja2022b}, propargyl cation \citep{Silva2023}, and propargyl radical \citep{Agundez2021}, one of the most abundant radicals found, have been discovered in TMC-1.
The study of this source is of extreme importance for the correct 
understanding of chemical processes in the interstellar medium. The results
obtained with these surveys also require
an improvement of the chemical networks as several 
of these molecules are not predicted by the current state-of-the-art 
chemical models.

Pure hydrocarbons have a crucial role in dark cloud reactions. Their weak rotational transitions, due to the low dipole moment, have complicated their study. However, thanks to the high sensitivity of the QUIJOTE data, we have been able to detect a large number of them, which 
will provide a census of pure hydrocarbons in TMC-1. The study of ethynyl and 
cyanide derivatives is of considerable importance as they allow 
us to follow the different chemical paths to producing them from the same 
precursor via reactions involving CCH and CN radicals.

In this Letter we report the detection of HCCCH$_2$CCH, with a total of 
18 lines detected with a S/N greater than 3$\sigma$ in the frequency 
range 31-50 GHz due to the high sensitivity of the survey. We also report an analysis of the five lowest-energy isomers with a non-zero dipole moment of the C$_5$H$_4$ family via a 
computational study, and we compare the abundance ratios of the detected species 
with their cyano derivative.

\section{Observations}

The data utilized in this study are part of the QUIJOTE spectral line 
survey of TMC-1 in the Q band, which was conducted with the Yebes 40m 
telescope at the position $\alpha_{J2000}=4^{\rm h} 41^{\rm m} 41.9^{\rm s}$ 
and $\delta_{J2000}=+25^\circ 41' 27.0''$. The receiver, developed under 
the Nanocosmos project, consists of two cold high-electron mobility transistor 
amplifiers that cover the 31.0--50.3 GHz band with horizontal and vertical 
polarizations. The backends are $2\times8\times2.5$ GHz fast Fourier transform 
spectrometers with a spectral resolution of 38 kHz, providing the whole 
coverage of the Q band in both polarizations. A description of the 
telescope, receivers, and backends is given by \citet{Tercero2021}.

\begin{figure*}
\centering
\includegraphics[width=0.9\textwidth]{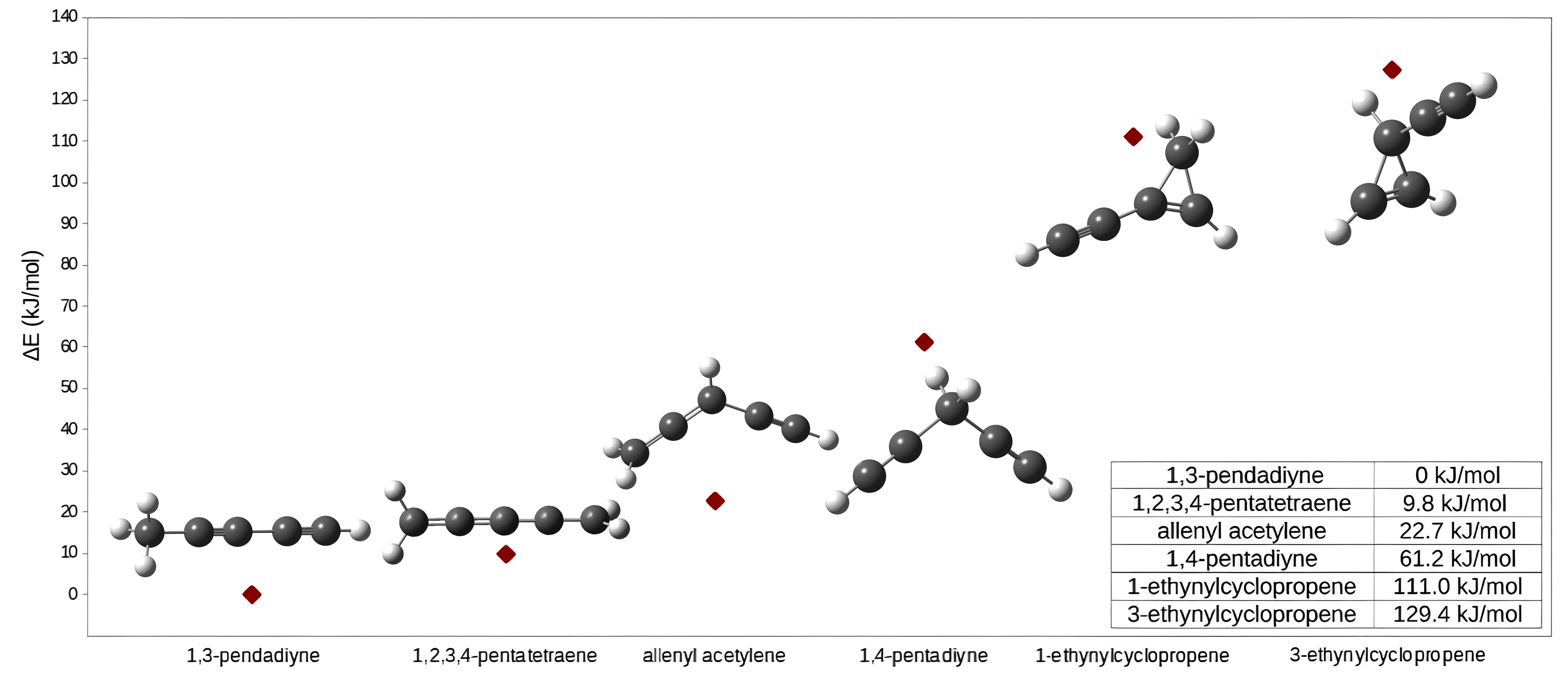} 
\caption{Relative energy of C$_5$H$_4$ isomers calculated at DFT(B3LYP)/cc-pVTZ.}\label{energy}
\end{figure*}

The QUIJOTE line survey was performed in multiple observing runs, and frequency-switching observing mode 
with a frequency throw of 8 and 10 MHz was used for all observations. The total observing time on the source for data taken 
with frequency throws of 8 MHz and 10 MHz (465 and 737 hours, respectively) was 1202 hours by November 2023. The noise level ranged from 0.09 mK at 32 GHz to 0.25 mK at 49.5 GHz, which 
is more than 50 times better than that of previous line surveys of TMC-1 in the Q band  \citep{Kaifu2004}. Different local 
oscillator frequencies were used for each frequency throw to remove any possible sideband effects in the down conversion chain. 
A detailed account of the QUIJOTE line survey is presented in \citet{Cernicharo2021a}.

The main beam efficiency varied from 0.66 at 32 GHz to 0.49 at 50 GHz \citep{Tercero2021}. The telescope beam size is
54.4$''$ and 36.4$''$ at 31 and 50 GHz, respectively. The intensity scale utilized in this study was antenna temperature 
($T_A^*$), calibrated using two absorbers at different temperatures, and the atmospheric transmission model \citep[][]{Cernicharo1985, Pardo2001}. Calibration uncertainties were adopted to be 10$\%$. All the data 
were analyzed with the GILDAS package\footnote{\texttt{http://www.iram.fr/IRAMFR/GILDAS}}.

\section{Results}

The C$_5$H$_4$ family has a large number of possible isomers. For the search for new isomers, 
\citet{Lattelais2009, Lattelais2010} proposed the principle of minimum energy as a method for obtaining new candidates, since most 
of the isomers detected are those of minimum energy. For this purpose, we performed geometry optimization calculations to obtain the 
most stable conformations using the Gaussian16 program package \citep{Frisch2016} and the B3LYP \citep{Lee1988} hybrid density functional with 
the Dunning’s correlation consistent polarized valence triple-($\zeta$) (cc-pVTZ) \citep{Woon1995}. The results are shown in 
Fig. \ref{energy} as a function of the energy difference (including zero-point energy) of the most stable isomer, 
CH$_3$C$_4$H. All the lower-energy molecules have already been detected (except 1,2,3,4-pentatetraene because it has 
no dipole moment), leaving 1-ethynylcyclopropene as the next candidate to be discovered. \citet{Kenny2001} performed theoretical calculations at different levels for several of these isomers. The values they obtained with the B3LYP functional and the DZP basis sets are quite similar to those derived for this work. The difference between the data derived by \citet{Kenny2001} and our estimates are 8.4 kJ mol$^{-1}$ for the two lowest-energy isomers and 7.6 kJ mol$^{-1}$ for 3-ethynylcyclopropene. However, there is a 
significant difference between our calculations of the coupled cluster, with differences amounting to 33.5 kJ mol$^{-1}$ and 16.8 kJ mol$^{-1}$, respectively.

Methyldiacetylene (CH$_3$C$_4$H) was the first C$_5$H$_4$ isomer detected in the interstellar medium \citep{Macleod1984,Walmsley1984}. The 
next detected isomer (CH$_2$CCHCCH) has an energy 22.7 kJ mol$^{-1}$ higher than the more stable species and was discovered by 
\citet{Cernicharo2021f}. The species HCCCH$_2$CCH is the only remaining non-cyclic isomer that can be detected, because the isomer 
H$_2$CCCCCH$_2$ is non-polar and therefore has no rotational lines. There are two additional cyclic isomers, which have a significantly 
higher energy with respect to the non-cyclic ones due to the presence of a cycle of three carbon atoms; this increases their energy relative 
to the other isomers. No spectroscopic information is available for these two isomers.

The line identification was achieved using the MADEX \citep{Cernicharo2012} and CDMS \citep{Muller2005} catalogues.
Rotational spectroscopy and the dipole moment ($\mu_b$ = 0.516D) for this species were measured in the 
laboratory by \citet{Kuczkowski1981}. 

\begin{table}
\tiny
\caption{Observed line parameters of HCCCH$_2$CCH.} \label{obs_line_parameters}
\resizebox{9cm}{!}{
\centering
\begin{tabular}{cccccl}
\hline
Transition$^a$     &$\nu_{rest}$~$^b$    & $\int T_A^* dv$~$^c$&  \multicolumn{1}{c}{$\Delta$v$^d$}    & \multicolumn{1}{c}{$T_A^*$~$^e$} & Notes  \\
                   &  (MHz)              & (mK\,km\,s$^{-1}$)  &  \multicolumn{1}{c}{(km\,s$^{-1}$)}    & \multicolumn{1}{c}{(mK)}       &  \\
\hline
3$_{1,3}$-2$_{0,2}$ &  31513.681$\pm$     0.015 &     0.57$\pm$ 0.16 &   1.07$\pm$  0.34 &     0.50$\pm$0.11 \\
4$_{1,4}$-3$_{0,3}$ &  36228.011$\pm$     0.010 &     0.40$\pm$ 0.14 &   0.71$\pm$  0.18 &     0.54$\pm$0.10 & A\\
9$_{0,9}$-8$_{1,8}$ &  36666.397 & & & & B \\
5$_{1,5}$-4$_{0,4}$&  40792.476$\pm$     0.010 &     0.76$\pm$ 0.13 &   1.11$\pm$  0.26 &     0.65$\pm$0.10 \\
12$_{2,10}$-12$_{1,11}$ &  41502.013$\pm$     0.013 &     0.27$\pm$ 0.08 &   0.58$\pm$  0.18 &     0.44$\pm$0.12 \\
11$_{2,9}$-11$_{1,10}$ &  42001.987$\pm$     0.019 &     0.18$\pm$ 0.10 &   0.43$\pm$  0.24 &     0.41$\pm$0.12 & C\\
10$_{2,8}$-10$_{1,9}$ &  42659.467$\pm$     0.010 &     0.39$\pm$ 0.07 &   0.56$\pm$  0.11 &     0.66$\pm$0.09 \\
10$_{0,10}$-9$_{1,9}$ &  42887.885$\pm$     0.010 &     0.36$\pm$ 0.05 &   0.50$\pm$  0.12 &     0.68$\pm$0.10 \\
9$_{2,7}$-9$_{1,8}$ &  43432.014$\pm$     0.010 &     0.21$\pm$ 0.09 &   0.43$\pm$  0.19 &     0.50$\pm$0.11 \\
8$_{2,6}$-8$_{1,7}$ &  44276.266$\pm$     0.010 &     0.67$\pm$ 0.11 &   0.72$\pm$  0.14 &     0.88$\pm$0.19 \\
7$_{2,5}$-7$_{1,6}$ &  45149.193$\pm$     0.010 &     0.20$\pm$ 0.05 &   0.38$\pm$  0.10 &     0.50$\pm$0.10 \\
6$_{1,6}$-5$_{0,5}$ &  45221.835$\pm$     0.010 &     0.85$\pm$ 0.13 &   0.81$\pm$  0.15 &     0.98$\pm$0.14 \\
6$_{2,4}$-6$_{1,5}$ &  46009.868$\pm$     0.010 &     0.60$\pm$ 0.10 &   0.73$\pm$  0.15 &     0.77$\pm$0.18  \\
5$_{2,3}$-5$_{1,4}$ &  46820.668$\pm$     0.010 &     0.60$\pm$ 0.20 &   0.69$\pm$  0.22 &     0.82$\pm$0.27  & C\\
4$_{2,2}$-4$_{1,3}$ &  47548.326$\pm$     0.010 &     0.35$\pm$ 0.06 &   0.38$\pm$  0.13 &     0.87$\pm$0.18 \\
3$_{2,1}$-3$_{1,2}$ &  48164.599$\pm$     0.020 &     0.36$\pm$ 0.12 &   0.57$\pm$  0.23 &     0.59$\pm$0.19 \\
11$_{0,11}$-10$_{1,10}$ &  49081.189$\pm$     0.018 &     0.69$\pm$ 0.19 &   0.76$\pm$  0.21 &     0.85$\pm$0.15 \\
7$_{1,7}$-6$_{0,6}$ &  49534.521$\pm$     0.010 &     0.72$\pm$ 0.22 &   0.58$\pm$  0.19 &     1.17$\pm$0.33 \\

\hline
\end{tabular}
}
\tablefoot{\\
\tablefoottext{a}{Quantum numbers are $J'_{K'_{a,}K'_{c}}$ - $J_{K_{a,}K_{c}}$.}\tablefoottext{b}{Observed frequency of the transition assuming a local standard of rest velocity of 5.83 km s$^{-1}$.}\tablefoottext{c}{Integrated line intensity in mK\,km\,s$^{-1}$.}\tablefoottext{d}{Linewidth at half intensity derived by fitting a Gaussian function to
the observed line profile in km\,s$^{-1}$.}\tablefoottext{e}{Antenna temperature in millikelvins.}\tablefoottext{A}{Frequency-switching data with a throw of 8 MHz only. Negative feature present in the data with a 10 MHz throw.}\tablefoottext{B}{This line is blended with a transition of HC$_4$N.}\tablefoottext{C}{Frequency-switching data with a throw of 10 MHz only. Negative feature present in the data with a 8 MHz throw.}
}
\end{table}

\begin{figure*} 
\centering
\includegraphics[width=0.75\textwidth]{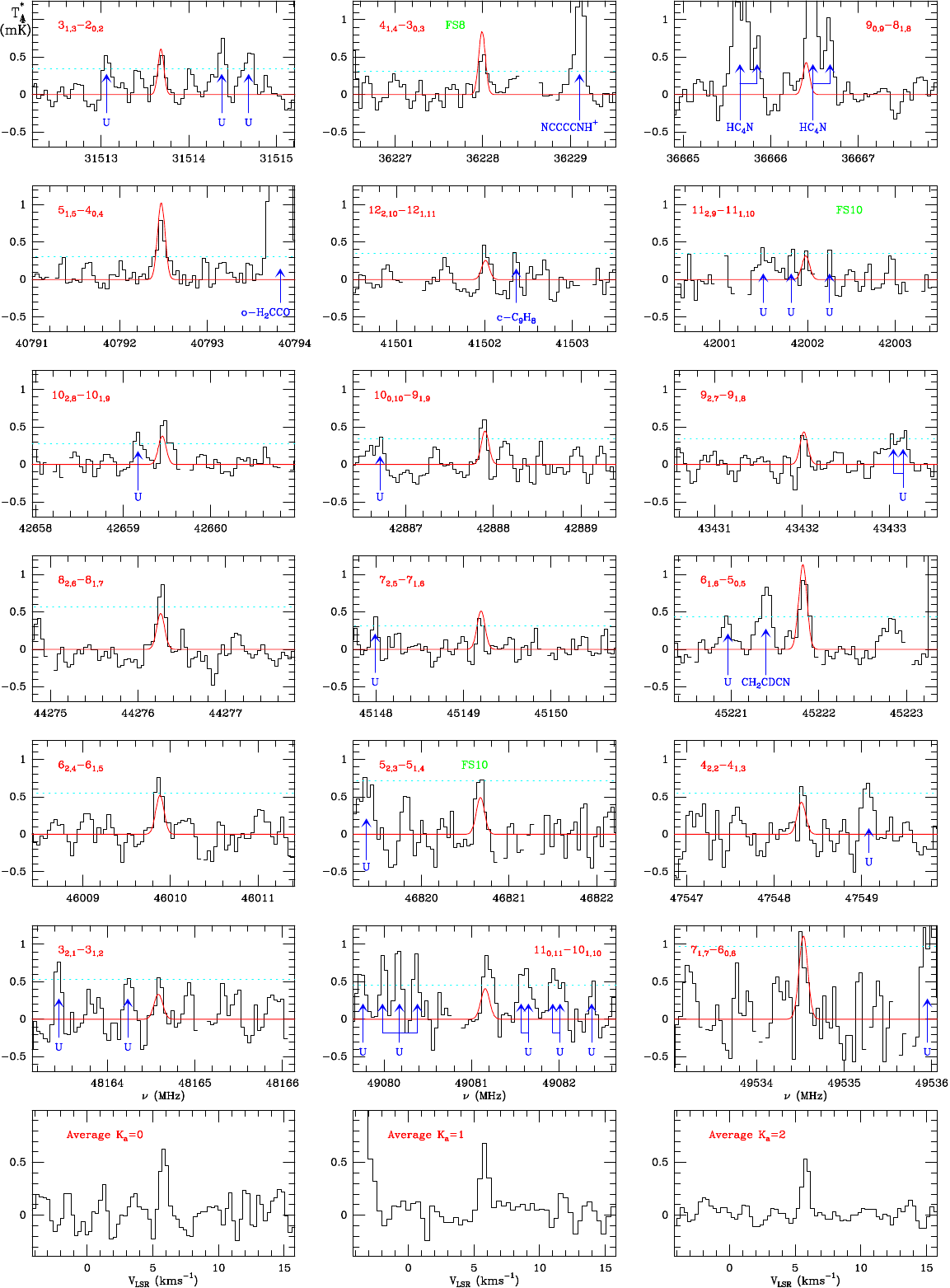} 
\caption{Selected transitions of HCCCH$_2$CCH in TMC-1.
The abscissa corresponds to the rest frequency of the lines. Frequencies and intensities for all observed lines
are given in Table \ref{obs_line_parameters}.
The ordinate is the antenna temperature, corrected for atmospheric and telescope losses, in millikelvins.
The quantum numbers of each transition are indicated
in the corresponding panel. The red line shows the computed synthetic spectra for this species for T$_{\mathrm{rot}}$ = 9.5 $\pm$ 0.5 K and
a column density of (5.0 $\pm$ 0.5)$\times$ 10$^{12}$ cm$^{-2}$ (see the main text). Blanked channels correspond to negative features 
produced when folding the frequency-switched data. Green labels indicate the transitions for which only one of the frequency-switching data points has been used (FS10
corresponds to a throw of 10 MHz). The three panels at the bottom show the average of all $K_a$=0,1, and 2 transitions. The noise level is below 0.1 mK for
these stacked spectra.
CH$_2$CDCN is reported in the $6_{16}-5_{05}$ transition panel. 
A detailed analysis of the isotopologues of CH$_2$CHCN will be presented elsewhere (Cernicharo et al., in prep).
}\label{lines}
\end{figure*}

We detected a total of 18 lines within the Q band, with peak intensities ranging from 1.17 mk to 0.41 mK, corresponding to the transitions from $J$ = 2 to $J$ = 13, with $K_a$ = 0, 1, and 2. 
The molecule has ortho and para species, so the spin statistic (5/3 for $K_a$ odd/even) was included in the calculations.
The derived line parameters for the molecules studied in this work were obtained by fitting a Gaussian line profile to the observed data (see Table \ref{obs_line_parameters}). A window of $\pm$ 15 \kms\, around the v$_{LSR}$ of the source was considered for each transition. 
Results in three of the lines were obtained only with the FSW data with a throw of 8 MHz or 10 MHz, since in the other one the line is affected by the FSW negative features (see notes in Table \ref{obs_line_parameters}, and Fig. \ref{lines}). The line  9$_{0,9}$-8$_{1,8}$ is blended with a transition of HC$_4$N.

To obtain the column density, we used a model line fitting procedure, with the local thermodynamic equilibrium approach for the thin optical lines (see e.g. \citealt{Cernicharo2021d}). We obtained $N$(HCCCH$_2$CCH) = (5.0$\pm$0.5$)\times$10$^{12}$ cm$^{-2}$ with a rotational temperature of 9.5$\pm$0.5 K. We assumed a source of uniform brightness with a diameter of 80$''$ \citep{Fosse2001}. The H$_2$ column density for TMC-1 is 10$^{22}$ cm$^{-2}$ (\citealp{Cernicharo1987}, Fuentetaja et al. in prep), so the abundance of HCCCH$_2$CCH is (5.0$\pm$0.5)$\times$10$^{-10}$. The predicted synthetic lines for these data are shown in Fig. \ref{lines}.

\begin{table}
\caption{Derived rotational temperatures and column densities for C$_5$H$_4$ and C$_4$H$_3$N isomers towards TMC-1.} \label{cd_isomers}
\centering
\begin{tabular}{ccc}
\hline
Molecule & T$_{rot}$ (K) & $N$ (cm$^{-2}$)\\
\hline
HCCCH$_2$CN $^{a}$ & 4 $\pm$ 1 & 2.8 $\pm$ 0.7 × 10$^{12}$ \\
CH$_2$CCHCN $^{a}$ &5.5 $\pm$ 0.3 & 2.7 $\pm$ 0.2 × 10$^{12}$ \\ 
$A$ - CH$_3$C$_3$N $^{a}$ & 6.7 $\pm$ 0.2 & 9.7 $\pm$ 0.3 × 10$^{11}$ \\
$E$ - CH$_3$C$_3$N $^{a}$ & 8.2 $\pm$ 0.6 & 7.7 $\pm$ 0.5 × 10$^{11}$ \\
CH$_2$CCHCCH $^{b}$& 7 $\pm$ 1 & 1.2 $\pm$ 0.2 × 10$^{13}$ \\
$A/E$ - CH$_3$C$_4$H $^{b}$&  7.0 $\pm$ 0.3 & 6.5 $\pm$ 0.3 × 10$^{12}$ \\
HCCCH$_2$CCH & 9.5 $\pm$ 0.5 & 5.0 $\pm$ 0.5 × 10$^{12}$\\
\hline
\end{tabular}
\tablefoot{\\
\tablefoottext{a}{Values reported by \citet{Marcelino2021}}\\
\tablefoottext{b}{Values reported by \citet{Cernicharo2021f}}\\
}
\end{table}

Assuming a column density of H$_2$ of 10$^{22}$ cm$^{-2}$ \citep{Cernicharo1987}, we see that the detected isomers of C$_5$H$_4$ show high abundances, ($\sim$2.5-12) $\times$ 10$^{-10}$.  It is interesting to analyse the abundance ratios of these molecules by substituting an ethynyl group for a cyano group (C$_4$H$_3$N) because the formation reactions occur through the same precursors, reacting with the CCH and CN radicals, respectively. It is observed that the cyano derivatives have slightly lower column densities (see Table \ref{cd_isomers}). All the abundance ratios obtained for these species are very similar. For the allenyl acetylene and cyano allene (CH$_2$CCHCCH/CH$_2$CCHCN), we obtain a value of 4.5, which was also reported by \citet{Cernicharo2021f}. For the most abundant isomers (CH$_3$C$_4$H/CH$_3$C$_3$N), we obtain a value of 3.7. Finally, from the column density obtained in this Letter, the abundance ratio HCCCH$_2$CCH/HCCCH$_2$CN = 1.8.

\section{Chemical model}

To study the chemistry of C$_5$H$_4$ isomers in TMC-1, we carried out calculations using a pseudo-time-dependent gas-phase chemical model. The model is based on the model presented by \cite{Cernicharo2021f} and is in line with the detection of CH$_2$CCHCCH in TMC-1. Briefly, we considered a gas kinetic temperature of 10 K, a volume density of H$_2$ of 2 \,$\times$\,10$^4$ cm$^{-3}$, a visual extinction of 30 mag, a cosmic-ray ionization rate of H$_2$ of 1.3\,$\times$\,10$^{-17}$ s$^{-1}$, and the so-called low-metal elemental abundances \citep{Agundez2013}. 

\begin{figure} 
\centering
\includegraphics[width=\columnwidth]{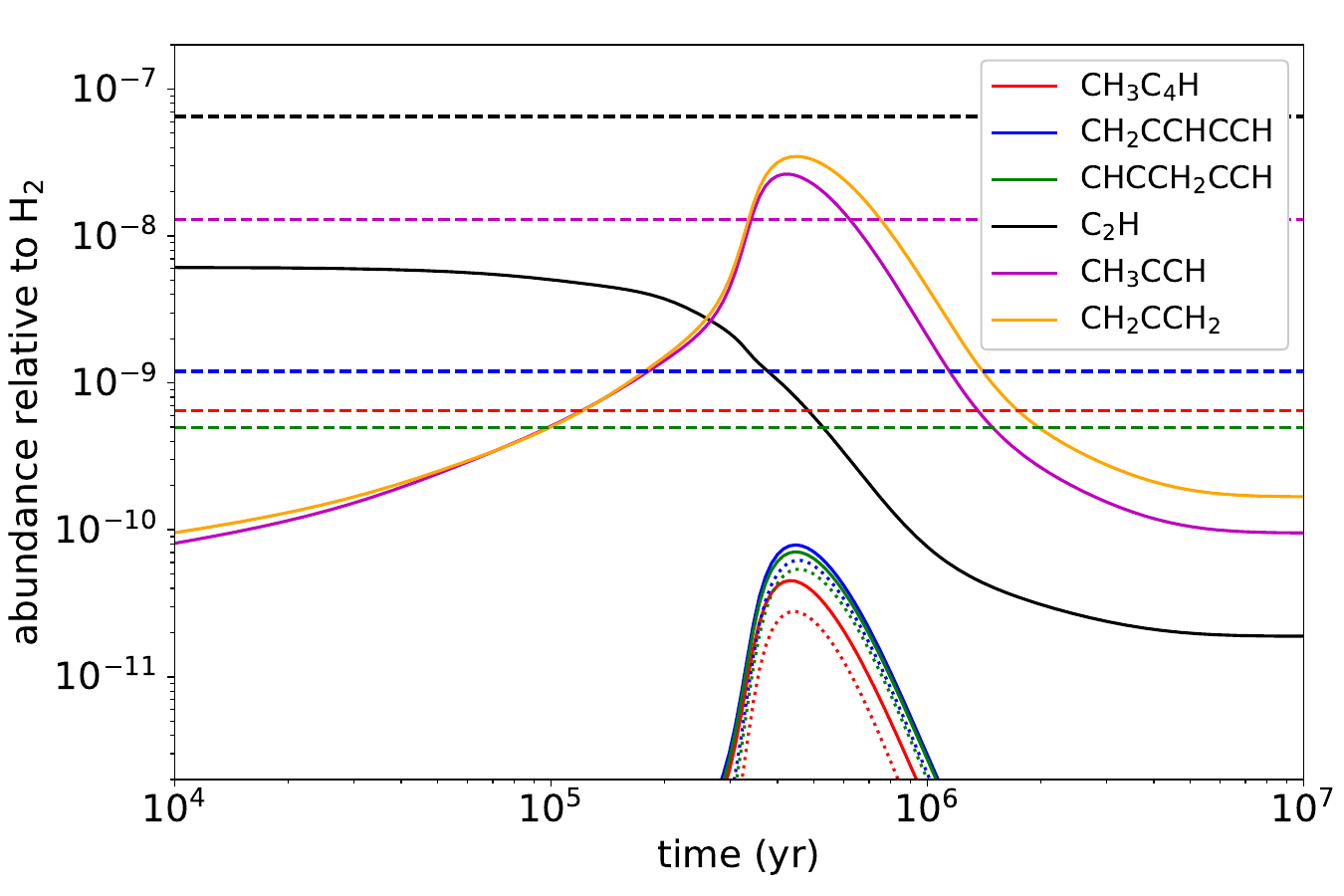} 
\caption{Fractional abundances of three C$_5$H$_4$ isomers calculated with the chemical model as a function of 
time. Dotted lines lines correspond to calculated abundances when the dissociative recombination of C$_5$H$_5^+$ 
is neglected as a source of C$_5$H$_4$ isomers. The abundances observed in TMC-1 (from this work 
and \citealt{Cernicharo2022}) are indicated by horizontal dashed lines.} \label{fig:abun}
\end{figure}

The core of the chemical network is based on RATE12 from the UMIST database \citep{McElroy2013}, with some updates (see \citealt{Marcelino2021}). We included three new C$_5$H$_4$ isomers, namely CH$_3$C$_4$H, CH$_2$CCHCCH, and CHCCH$_2$CCH. The formation and destruction of the three isomers are based on the reactions previously considered for CH$_3$C$_4$H, which is the only C$_5$H$_4$ isomer considered in the UMIST database. In particular, the three isomers are destroyed by reactions with C neutral atoms with abundant cations, such as C$^+$, H$^+$, H$_3^+$, HCO$^+$, and He$^+$, and with secondary ultraviolet photons arising from cosmic ray impacts. Among the reactions of formation of CH$_3$C$_4$H included in the UMIST database, there is the dissociative recombination of C$_5$H$_5^+$, although there is no evidence as to whether C$_5$H$_4$ isomers are produced in this reaction. In the absence of better constraints, we assumed that the three C$_5$H$_4$ isomers are produced with equal yields. The UMIST database also includes the reaction between C$_2$ and propene as a source of CH$_3$C$_4$H. This reaction is known to be rapid at low temperatures \citep{Daugey2008}, although the main products are C$_5$H$_5$ isomers, rather than C$_5$H$_4$ isomers \citep{Dangi2013}. We therefore did not include this reaction as a source of C$_5$H$_4$ isomers. The remaining processes of CH$_3$C$_4$H  formation included in the UMIST database are the reactions of C$_2$H with methylacetylene and allene. These reactions have been widely studied through experiments and theoretical calculations. First, the reactions were measured to be fast at low temperatures \citep{Hoobler1999,Vakhtin2001,Carty2001}. The rate coefficient measured at 63 K is 2.9\,$\times$\,10$^{-10}$ cm$^3$ s$^{-1}$ for C$_2$H + CH$_3$CCH and 3.5\,$\times$\,10$^{-10}$ cm$^3$ s$^{-1}$ for C$_2$H + CH$_2$CCH$_2$ \citep{Carty2001}. The product distribution has also been studied through crossed beams experiments \citep{Kaiser2001,Zhang2009}, ab initio calculations \citep{Stahl2001,Jamal2010}, and experiments using mass spectrometry and synchrotron photoionization \citep{Goulay2007,Goulay2011}. These studies indicate that the reaction of C$_2$H with methylacetylene produces diacetylene and the C$_5$H$_4$ isomers CH$_3$C$_4$H and CH$_2$CCHCCH \citep{Goulay2007,Jamal2010}, while the reaction of C$_2$H with allene produces the C$_5$H$_4$ isomers CH$_2$CCHCCH and CHCCH$_2$CCH \citep{Goulay2007,Zhang2009,Jamal2010}. We thus included the reactions
\begin{subequations} \label{reac:c2h+ch3cch}
\begin{align}
\rm C_2H + CH_3CCH & \rightarrow \rm C_4H_2 + CH_3 & 50\,\%, \label{reac:c2h+ch3cch_a} \\
                                   & \rightarrow \rm CH_3C_4H + H & 40\,\%, \label{reac:c2h+ch3cch_b} \\
                                   & \rightarrow \rm CH_2CCHCCH + H & 10\,\%, \label{reac:c2h+ch3cch_c}
\end{align}
\end{subequations}
\begin{subequations} \label{reac:c2h+ch2cch2}
\begin{align}
\rm C_2H + CH_2CCH_2 & \rightarrow \rm CH_2CCHCCH + H & 50\,\%, \label{reac:c2h+ch2cch2_a} \\
                                     & \rightarrow \rm CHCCH_2CCH + H & 50\,\%, \label{reac:c2h+ch2cch2_b}
\end{align}
\end{subequations}
where the branching ratios are based on the aforementioned studies. We adopted the rate coefficients measured at 63 K \citep{Carty2001}. We note that the reaction of C$_2$H with allene has also been found to yield the non-polar C$_5$H$_4$ isomer CH$_2$CCCCH$_2$, although the yield should be lower than 10\,\% \citep{Goulay2011}. The C$_5$H$_4$ isomer ethynylallene (CH$_2$CCHCCH) is expected to be the major product of the reaction CH + CH$_2$CHCCH \citep{He2022}, and thus we included this reaction with a rate coefficient of 3\,$\times$\,10$^{-10}$ cm$^3$ s$^{-1}$, which is of the order of the values measured at low temperatures for many other rapid neutral-neutral reactions.

The resulting abundances of the three isomers of C$_5$H$_4$ detected in TMC-1 are shown in Fig.\,\ref{fig:abun}. The peak abundances of the three C$_5$H$_4$ isomers, reached at a time of (4-5)\,$\times$\,10$^5$ yr, are very similar. At these times, the main reactions of formation of the three C$_5$H$_4$ isomers are the reactions of C$_2$H with CH$_3$CCH and CH$_2$CCH$_2$, together with the dissociative recombination of C$_5$H$_5^+$. The dotted lines in Fig. 4 show the calculated abundances when the latter pathway, which is more speculative, is neglected. Since the precursors methylacetylene and allene are predicted to have similar abundances (allene being slightly more abundant than methylacetylene; see Fig.\,\ref{fig:abun}) and since the branching ratios yielding the three isomers in reactions (\ref{reac:c2h+ch3cch}) and (\ref{reac:c2h+ch2cch2}) are also similar, the resulting abundances are very close. This is in agreement with the fact that observed abundances are also quite similar. The main discrepancy between the model and the observations is that the calculated abundances lie one order of magnitude below the observed ones. In Fig.\,\ref{fig:abun} we also plot the abundance of the precursors involved in reactions (\ref{reac:c2h+ch3cch}) and (\ref{reac:c2h+ch2cch2}). While the calculated abundance of CH$_3$CCH is of the order of the observed value, the model underestimates the observed abundance in the case of C$_2$H, which translates into the underestimation of the abundances of the three C$_5$H$_4$ isomers. The lesson learnt from the chemical model is that the reactions of C$_2$H with methylacetylene and allene are the main sources of C$_5$H$_4$ isomers in TMC-1. The underestimation of the abundance of HCCCH$_2$CCH is likely due to an underestimation of the abundance of the precursor C$_2$H. Alternatively, there could be other formation routes missing in the chemical model.

\section{Conclusions}

We report the first detection of HCCCH$_2$CCH in TMC-1. We observed a total of 18 rotational transitions, from $J$ = 2 to 13 and $K_a\le 2$, using the Yebes 40m telescope. The rotational temperature obtained is 9.5$\pm$0.5 K and the derived column densities are (5.0$\pm$0.5)$\times$10$^{12}$ cm$^{-2}$.  We carried out a comparison of the energies of their five most stable isomers using theoretical calculations, and we obtained different abundance ratios with their cyano derivatives in order to contribute to the understanding of the chemical processes involving these species. The abundances of the three isomers are very similar, which is in agreement with the branching yielding ratios of the reactions of C$_2$H with methylacetylene and allene, although the model underestimates their values.

\begin{acknowledgements}
We thank Ministerio de Ciencia e Innovaci\'on of Spain (MICIU) for funding support through projects
PID2019-106110GB-I00, PID2019-107115GB-C21 / AEI / 10.13039/501100011033, and
PID2022-137980NB-100. We also thank ERC for funding
through grant ERC-2013-Syg-610256-NANOCOSMOS. M.A. thanks MCIU for grant RyC-2014-16277.
\end{acknowledgements}

\normalsize


\begin{thebibliography}{} 
\tiny
\bibitem[Ag\'undez \& Wakelam(2013)]{Agundez2013} Ag\'undez, M. \& Wakelam, V. 2013, Chem. Rev., 113, 8710
\bibitem[Ag\'undez et al. (2021)]{Agundez2021} Ag\'undez, M., Cabezas, C., Tercero, B., et al. 2021, \aap, 647, L10 
\bibitem[Carty et al. (2001)]{Carty2001} Carty, D., Le Page, V., Sims, I. R., \& Smith, I. W. M. 2001, Chem. Phys. Lett., 344, 310
\bibitem[Cernicharo(1985)]{Cernicharo1985} Cernicharo, J. 1985, Internal IRAM report (Granada: IRAM)
\bibitem[Cernicharo \& Gu\'elin(1987)]{Cernicharo1987} Cernicharo, J. \& Gu\'elin, M. 1987, \aap, 176, 299
\bibitem[Cernicharo (2012)]{Cernicharo2012} Cernicharo, J., 2012, in ECLA 2011: Proc. of the European Conference on Laboratory Astrophysics,
EAS Publications Series, 2012, Ed.: C. Stehl, C. Joblin, \& L. d'Hendecourt (Cambridge: Cambridge Univ. Press),
251; \texttt{https://nanocosmos.iff.csic.es/?page$\_$id=1619}
\bibitem[Cernicharo et al.(2021a)]{Cernicharo2021a} Cernicharo, J., Ag\'undez, M., Kaiser, R., et al. 2021a, \aap, 652, L9 
\bibitem[Cernicharo et al.(2021b)]{Cernicharo2021b} Cernicharo, J., Ag\'undez, M., Cabezas, C., et al. 2021b, \aap, 647, L2  
\bibitem[Cernicharo et al.(2021c)]{Cernicharo2021c} Cernicharo, J., Ag\'undez, M., Cabezas, C., et al.  2021c, \aap, 649, L15 
\bibitem[Cernicharo et al.(2021d)]{Cernicharo2021d} Cernicharo, J., Cabezas, C., Endo, Y., et al. 2021d, \aap, 646, L3 
\bibitem[Cernicharo et al.(2021e)]{Cernicharo2021e} Cernicharo, J., Cabezas, C., Ag\'undez, M., et al. 2021e, \aap, 648, L3 
\bibitem[Cernicharo et al.(2021f)]{Cernicharo2021f} Cernicharo, J., Cabezas, C., Ag{\'u}ndez, M., et al. 2021f, \aap, 647, L3 
\bibitem[Cernicharo et al.(2022)]{Cernicharo2022} Cernicharo, J., Fuentetaja, R., Ag\'undez, M., et al. 2022, \aap, 663, L9
\bibitem[Dangi et al.(2013)]{Dangi2013} Dangi, B.~B., Maity, S., Kaiser, R.~I., et al.\ 2013, Journal of Physical Chemistry A, 117, 11783
\bibitem[Daugey et al.(2008)]{Daugey2008} Daugey, N., Caubet, P., Bergeat, A., et al. 2008, Phys. Chem. Chem. Phys., 10, 729
\bibitem[Frisch et al.(2016)]{Frisch2016} Frisch, M. J., Trucks, G. W., Schlegel, H. B., et al. 2016, Gaussian 16 Revision A.03
\bibitem[Foss\'e et al. (2001)]{Fosse2001} Foss\'e, D., Cernicharo, J., Gerin, M., Cox, P. 2001, \apj, 552, 168
\bibitem[Fuentetaja et al. (2022a)]{Fuentetaja2022} Fuentetaja, R., Cabezas, C., Ag\'undez, M., et al. 2022a, \aap,  663, L3
\bibitem[Fuentetaja et al.(2022b)]{Fuentetaja2022b} Fuentetaja, R., Ag{\'u}ndez, M., Cabezas, C., et al.\ 2022b, \aap, 667, L4
\bibitem[Goulay et al.(2007)]{Goulay2007} Goulay, F., Osborn, D. L., Taatjes, C. A., et al. 2007, Phys. Chem. Chem. Phys., 9, 4291
\bibitem[Goulay et al.(2011)]{Goulay2011} Goulay, F., Soorkia, S., Meloni, G., et al. 2011, Phys. Chem. Chem. Phys., 13, 20820
\bibitem[He et al.(2022)]{He2022} He, C., Yang, Z., Doddipatla, S., et al. 2022, Phys. Chem. Chem. Phys., 24, 26499
\bibitem[Hoobler \& Leone(1999)]{Hoobler1999} Hoobler, R. J. \& Leone, S. R. 1999, J. Phys. Chem. A, 103, 1342
\bibitem[Jamal \& Mebel(2010)]{Jamal2010} Jamal, A. \& Mebel, A. M. 2010, Phys. Chem. Chem. Phys., 12, 2606
\bibitem[Kaifu et al.(2004)]{Kaifu2004} Kaifu, N., Ohishi, M., Kawaguchi, K., et al. 2004, PASJ, 56, 69
\bibitem[Kaiser et al.(2001)]{Kaiser2001} Kaiser, R. I., Chiong, C. C., Asvany, O., et al. 2001, Chem. Phys. 114, 3488
\bibitem[Kenny et al.(2001)]{Kenny2001} Kenny, J.~P., Krueger, K.~M., Rienstra-Kiracofe, J.~C., et al.\ 2001, Journal of Physical Chemistry A, 105, 7745
\bibitem[Kuczkowski et al.(1981)]{Kuczkowski1981} Kuczkowski, R.~L., Lovas, F.~J., Suenram, R.~D., et al.\ 1981, Journal of Molecular Structure, 72, 143
\bibitem[Lattelais et al.(2009)]{Lattelais2009} Lattelais, M., Pauzat, F., Ellinger, Y., et al.\ 2009, \apjl, 696, L133
\bibitem[Lattelais et al.(2010)]{Lattelais2010} Lattelais, M., Pauzat, F., Ellinger, Y., et al.\ 2010, \aap, 519, A30
\bibitem[Lee et al.(1988)]{Lee1988} Lee, C., Yang, W., \& Parr, R.~G.\ 1988, \prb, 37, 785
\bibitem[Lee et al.(2021)]{Lee2021} Lee, K.~L.~K., Changala, P.~B., Loomis, R.~A., et al.\ 2021, \apjl, 910, L2
\bibitem[Loru et al.(2023)]{Loru2023} Loru, D., Cabezas, C., Cernicharo, J., et al.\ 2023, \aap, 677, A166
\bibitem[MacLeod et al.(1984)]{Macleod1984} MacLeod, J.~M., Avery, L.~W., \& Broten, N.~W.\ 1984, \apjl, 282, L89
\bibitem[Marcelino et al.(2021)]{Marcelino2021} Marcelino, N., Tercero, B., Ag{\'u}ndez, M., et al.\ 2021, \aap, 646, L9 
\bibitem[McCarthy et al.(2021)]{McCarthy2021} McCarthy, M.~C., Lee, K.~L.~K., Loomis, R.~A., et al.\ 2021, Nature Astronomy, 5, 176
\bibitem[McElroy et al.(2013)]{McElroy2013} McElroy, D., Walsh, C., Markwick, A. J., et al. 2013, \aap, 550, A36
\bibitem[McGuire et al.(2018)]{McGuire2018} McGuire, B.A., Burkhardt, A.M., Kalenskii, S., et al. 2018, Science, 359, 202
\bibitem[McGuire et al.(2021)]{McGuire2021} McGuire, B.~A., Loomis, R.~A., Burkhardt, A.~M., et al.\ 2021, Science, 371, 1265
\bibitem[M\"uller et al.(2005)]{Muller2005} M\"uller, H.S.P., Schl\"oder, F., Stutzki, J., Winnewisser, G. 2005, \jmst, 742, 215
C. Lee, W. Yang and R. G. Parr, Phys. Rev. B: Condens. Matter Mater. Phys., 1988, 37, 785–789.

\bibitem[Pardo et al.(2001)]{Pardo2001} Pardo, J.~R., Cernicharo, J., Serabyn, E. 2001, IEEE Trans. Antennas and Propagation, 49, 12
\bibitem[Silva et al.(2023)]{Silva2023} Silva, W.~G.~D.~P., Cernicharo, J., Schlemmer, S., et al.\ 2023, \aap, 676, L1
\bibitem[Sita et al.(2022)]{Sita2022} Sita, M.~L., Changala, P.~B., Xue, C., et al.\ 2022, \apjl, 938, L12
\bibitem[Stahl et al.(2001)]{Stahl2001} Stahl, F., von Rahu\'e Schleyer, P., Bettinger, H. F., et al. 2001, J. Chem. Phys., 114, 3476
\bibitem[Tercero et al.(2021)]{Tercero2021} Tercero, F., L\'opez-P\'erez, J. A., Gallego, et al. 2021, \aap, 645, A37
\bibitem[Vakhtin et al.(2001)]{Vakhtin2001} Vakhtin, A. B., Heard, D. E., Smith, I. W. M., \& Leone, S. R. 2001, Chem. Phys. Lett., 344, 317
\bibitem[Walmsley et al.(1984)]{Walmsley1984} Walmsley, C.~M., Jewell, P.~R., Snyder, L.~E., et al.\ 1984, \aap, 134, L11
\bibitem[Woon \& Dunning(1995)]{Woon1995} Woon, D.~E. \& Dunning, T.~H.\ 1995, \jcp, 103, 4572
\bibitem[Zhang et al.(2009)]{Zhang2009} Zhang, F., Kim, S., \& Kaiser, R. I. 2009, Phys. Chem. Chem. Phys., 11, 4707

\end{thebibliography}
\end{document}